\documentclass[twocolumn]{aa}
\usepackage{graphicx,natbib,psfrag,amssymb,amsmath}
\usepackage{txfonts}
\usepackage{color}
\usepackage{textcomp}
\definecolor{noire}{rgb}{0,0,0} 
\definecolor{blue}{rgb}{0.1,0.5,0.7}
\bibpunct{(}{)}{;}{a}{}{,}

\usepackage{natbib}
\bibpunct{(}{)}{;}{a}{}{,}

\begin{document}

   \title{Highly-magnified, Multiply-imaged radio counterparts of the
Sub-mm Starburst Emission in the Cluster-Lens MS0451.6$-$0305}

   \subtitle{}
 
   \author{A. Berciano Alba\inst{1,2}, M.A. Garrett \inst{1},
   L.V.E. Koopmans\inst{2} \& O. Wucknitz\inst{1}}

   \offprints{berciano@astro.rug.nl}

   \institute{ Joint Institute for VLBI in Europe, Postbus 2, 7990 AA,
     Dwingeloo, The Netherlands \and Kapteyn Astronomical Institute,
     University of Groningen, P.O.box 800, 9700 AV, Groningen, The
     Netherlands}
 
   \date{Received ...; accepted ...}

   \abstract{Previous authors have reported the detection of
     intrinsically faint sub-mm emission lensed by the cluster
     MS0451.6$-$0305. They suggest that this emission arises from a
     merging system composed of a Ly-break galaxy and a pair of
     extremely red objects which are multiply-imaged in the
     optical/NIR observations.} {Since the submm emission presents an
     unusually large angular extent ($\sim$1\arcmin), the possible
     radio emission asociatted with that system can help to identify
     optical/NIR counterparts due to the higher spatial resolution and
     astrometric accuracy of the radio observations.}{Archive VLA data
     (BnA configuration at 1.4 GHz) was reduced and analysed. A simple
     lens model was constructed to aid the interpretation of the radio
     and pre-existing sub-mm and optical/NIR data.}{We present a 1.4
     GHz map of the central region of MS0451.6$-$0305 and report the
     detection of gravitationally lensed radio emission, coincident
     with the previously discovered sub-mm lensed emission. The
     overall morphology and scale of the radio and sub-mm emission are
     strikingly similar, extending $\sim 1\arcmin$ across the
     sky. This observation strongly suggests that the radio and sub-mm
     emission arise from the same sources. Preliminary estimates of
     the total $S_{850 \mu {\rm m}}/S_{1.4 {\rm GHz}}$ flux density
     ratio appear to be consistent with that expected from distant
     star forming galaxies. The radio emission is resolved into 7
     distinct components, and the overall structure can be explained,
     using a simple lens model, with three multiply-imaged radio
     sources at z $\sim2.9$. One of these sources is predicted to lie
     in the middle of the previously mentioned system in the source
     plane, suggesting that it is related to the intense star
     formation generated during the merging process.}{}
  
   \keywords{Gravitational Lensing - Galaxies: starburst - Radio
continuum:galaxies - Interacting group}

 \titlerunning{Radio counterpart to the sub-mm emission in MS0451.6$-$0305}
 \authorrunning{Berciano Alba et al.}

   \maketitle

\section{Introduction}

Sub-mm galaxies (SMGs) were first detected by
SCUBA\footnote{Submillimetre Common-User Bolometer Array, mounted at
the James Clerk Maxwell Telescope (JCMT)} \citep{SIB97} and are
believed to be dusty star forming galaxies located at high
redshift \citep{I02,S02,C03}. It is also suggested that they are the
progenitors of present-day massive elliptical galaxies
\citep[e.g.][]{lilly99, swinbank06}. Little is known about the objects
associated with the faint end of the SMG population ($S_{850 \mu
m}<2$mJy), but they are predicted to dominate (energetically) the
population as a whole \citep{KN04}. A recent statistical stacking
analysis \citep{KN05}, suggests that distant red galaxies (DRGs) and
Extremely Red Objects (EROs) contribute $\sim 50$\% of the flux
density of sub-mm sources with $0.5 < S_{850 \mu m}<5$mJy.

\begin{figure*}[ht!]
  \centering 

\includegraphics[width=8cm,angle=-90]{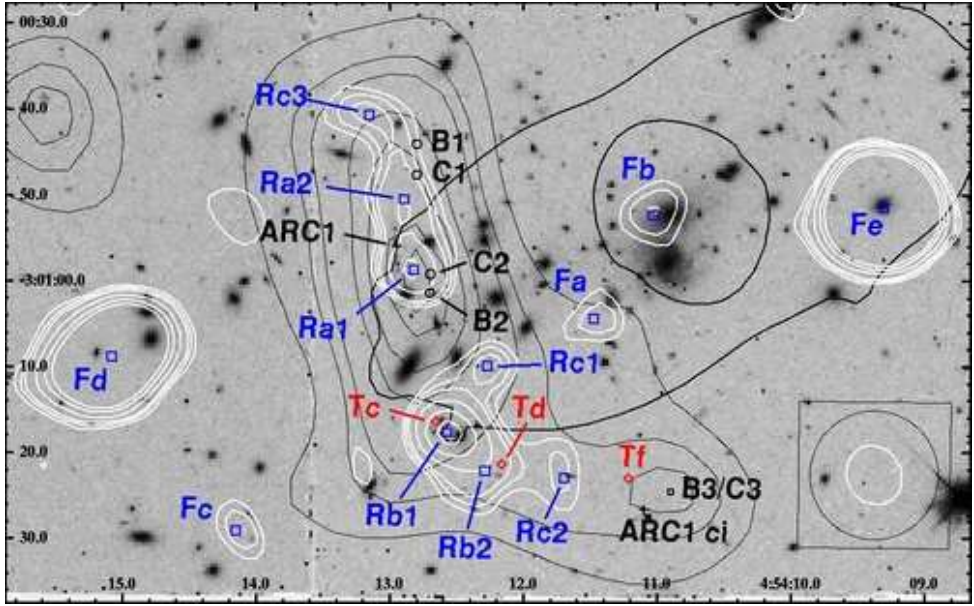}
\caption{The VLA 1.4 GHz naturally weighted contour map (solid white
lines) superimposed upon the SCUBA 850-$\mu$m contour map (solid thin
black lines) and the inverted HST F702W image of the centre of the
cluster MS0451.6$-$0305 \citep{B04}. The axes represent the right
ascension (x-axis) and declination (y-axis) in the J2000 coordinate
system. The solid thick black curves are the tangential (outer) and
radial (inner) critical lines at z=2.911 associated with the lens
model of the cluster determined by \cite{B04}. The boxes are the
positions obtained via Gaussian fits of the radio sources. The
diamonds are the positions of three EROs from \cite{T03}, and the
crosses/circles are the positions of a LBG lensed as two arc
(\emph{ARC1} and \emph{ARC1 ci}), and a triply-imaged EROs pair:
\emph{B1/B2/B3} (images of ERO B) and \emph{C1/C2/C3} (images of ERO
C) \cite[see][]{B04}. The squares and circles have a size of 1\arcsec\
to illustrate the random and systematic errors due to measurement
indeterminations and the aligment of the different images.  Contours
of the \emph{radio map} are drawn at -3, 3, 4, 5, 8, 12 $\&$ 16 times
the1-$\sigma$ noise level of 9 $\mu$Jy per beam. Contours of the
\emph{sub-mm map} are drawn at 4, 6, 7, 9, 10, 11 $\&$ 11.5 mJy per
beam. The white circle inside a box in the bottom-right corner is the
beam-size of the radio map (6.99 $\times$ 6.03 arcsec in position
angle $PA=32.6^{\circ}$) whereas the black one corresponds to the
beam-size of the sub-mm map (15 $\times$ 15 arcsec).}
  \label{natural}
  \end{figure*}

Intrinsically faint SMG cannot easily be detected, since their flux
densities lie below the $\sim 2$ mJy confusion limit of SCUBA images
at 850 $\mu$m. Typically they also fall well below the (thermally
limited) sensitivity of current radio instruments, such as the
VLA. Individual systems can often only be detected via strong
gravitational lensing effects, produced by massive foreground clusters
of galaxies \citep{KN04, K04, G05}.

The spatial magnification provided by the lensing cluster overcomes
instrumental confusion limitations in the sub-mm and also boosts the
measured flux density of the source (provided the lensed images remain
unresolved), thereby increasing the probability of detection
\citep{blain97}. Another advantage of cluster lensing is that the
magnification provided by the lens effectively increases the spatial
resolution of the observations, with the largest magnifications
usually occurring in cases of multiple imaging.

SMM~J16359+6612, associated with the cluster Abell 2218, was the first
intrinsically faint, multiply imaged SMG detected in both the sub-mm
\citep{K04} and radio \citep{G05}. In this paper, we present VLA 1.4
GHz radio observations of a second case, SMM~J04542$-$0301
\citep{C02}, associated with the cluster MS0451.6$-$0305.

MS0451.6$-$0305 is a cluster of galaxies situated at z=0.55
\citep{GL94} that has recently been studied using optical and
near-Infrared (NIR) data \citep{B04}. They conclude that the sub-mm
emission is probably related to an interacting system of three objects
lying at $z \sim 2.9$: a Lyman Break Galaxy (LBG) and a pair of
Extremely Red Objects (EROs). In the optical and NIR images, it is
proposed that the LBG is imaged into two visible arcs, and the ERO
pair are responsible for 5 additional sources of emission 
in the field \citep{B04, T03}. However, the emission coming from the
north-eastern and the central regions of the sub-mm image, cannot be fully 
reproduced using the LBG and the ERO pair alone \cite[see Fig.7
from][]{B04}.

In this paper, we present deep, high resolution 1.4 GHz VLA
observations of SMM~J04542$-$0301. In Sect.
\ref{sec:radio-observations}, we describe the VLA data analysis and
present the associated radio images. Section \ref{sec:lens-model}
describes a simple lens model for the system, in an attempt to explain
the lensed nature of the radio emission related to SMM
J04542$-$0301. In Sect. \ref{sec:comparing-radio-sub}, we compare the
radio and sub-mm emission, including a discussion about possible
optical/NIR counterparts, and the preliminary calculation of the
$S_{850 \mu {\rm m}}/S_{1.4 {\rm GHz}}$ flux density ratio. A summary
of our main results is presented in Sect. \ref{sec:conclusions}.
In the following discussion, we assume a $\Lambda$CDM cosmological model
with $\Omega_{m}$=0.3 , $\Omega_{\lambda}$=0.7 and $h_{0}$=0.7

\section{Radio Observations}  \label{sec:radio-observations}

\begin{table*}[ht!]
\centering
\caption{ \textbf{Details of the radio sources observed in the core of
 MS0451.6-0305}. The columns show: position (RA, DEC), peak flux
 density (S$_{Pk}$), total flux density (S$_{T}$) and deconvolved
 Gaussian sizes (major axis, minor axis and position angle) with their
 corresponding formal errors. Cases where the parameters of the
 Gaussian fits are not well constrained are indicated by a
 dash. The coordinates are given as offsets with respect to the
 cluster centre, RA(J2000)=04:54:10.8 and DEC(J2000)=-03:00:51.6
 \cite[see][table 2]{T03}. A version of this table in absolute
 coordinates can be found in the online material.}
\begin{tabular}{lccccccc}
\hline\hline\\ Name & RA & DEC & S$_{Pk}$ & S$_{T}$ & Maj Axis & Min Axis & PA \\ 
                    & J2000 (\arcsec) & J2000 ($\arcsec$)& $\mu$Jy & $\mu$Jy & ($\arcsec$) & ($\arcsec$) & deg
\\

\hline \\
Ra2 &     31.3$\pm$0.3 & $1.1\pm0.4$    & $70\pm8$   & $95\pm18$   & $6\pm1$      & $2\pm1$     & $27\pm28$  \\
Ra1 &     30.3$\pm$0.2 & $-7.1\pm0.2$   & $109\pm9$  & $109\pm9$   & $-$          & $-$         & $-$        \\
Rb1 &     26.5$\pm$0.2 & $-26\pm0.2$    & $151\pm9$  & $151\pm9$   & $-$          & $-$         & $-$        \\
Rb2 &     22.3$\pm$0.5 & $-30.50\pm0.7$ & $52\pm8$   & $100\pm22$  & $9\pm2$      & $3\pm2$     & $158\pm14$ \\
Rc1 &     22.1$\pm$0.6 & $-18.3\pm0.4$  & $50\pm9$   & $55\pm16$   & $6\pm2$      & $-$         & $112\pm11$ \\
Rc2 &     13.5$\pm$0.6 & $-31.3\pm0.8$  & $41\pm8$   & $58\pm18$   & $7\pm2$      & $1\pm3$     & $10\pm163$ \\
Rc3 &     35.4$\pm$0.8 & $10.9\pm0.5$   & $52\pm8$   & $78\pm19$   & $8\pm2$      & $0\pm2$     & $73\pm11$  \\

\hline

Fa &     10.1$\pm$0.6  & $-12.7\pm0.6$  & $45\pm9$   & $50\pm17$   & $4\pm6$      & $0\pm4$     & $123\pm38$ \\
Fb &     3.5 $\pm$0.6  & $-0.8\pm0.6$   & $49\pm9$   & $70\pm20$   & $6\pm5$      & $3\pm6$     & $122\pm45$ \\
Fc &     50.4$\pm$0.6  & $-37.5\pm0.6$  & $44\pm9$   & $44\pm9$    & $-$          & $-$         & $-$        \\
Fd &     64.1$\pm$0.1  & $-17.2\pm0.1$  & $634\pm9$  & $1039\pm21$ & $7.6\pm0.1$ & $1.4\pm0.5$ & $128\pm1$  \\
Fe & $-$22.40$\pm$0.02 & $-0.07\pm0.02$ & $1549\pm9$ & $1777\pm17$ & $3.3\pm0.1$  & $0.8\pm0.5$ & $125\pm3$  \\

\hline

\end{tabular}
\label{radio_data}
\end{table*}

VLA 1.4 GHz observations of the cluster MS0451.6$-$0305 were made in
June 2002, and were retrieved from the NRAO data archive
system\footnote{project ID AN0109, PI: Nakanishi}. The integration
time was 7.8 hours with the VLA in BnA configuration, employing two 25
MHz IFs in both left and right-hand circular polarization. Each IF
was subdivided into 7 channels. The data analysis was performed using
the NRAO AIPS package using standard analysis techniques. The absolute
flux density scale was set by observations of 0137+331, and phase
calibration was performed via short observations of 0503+020 between
the 1 hour target scans. A wide-field image was made and bright
sources far from the field centre were subtracted from the
data. Self-calibration using the remaining sources in the centre of
the field realised images with a 1$\sigma$ r.m.s. noise level of 9
$\mu$Jy/beam.

In Fig.\ref{natural}, we present the radio contour map (solid white
lines) of the naturally weighted VLA image of SMM J04542$-$0301,
superimposed upon the HST F702W image and the sub-mm contour map
(solid thin black lines) presented in \cite{B04}. Note that the SCUBA beam is
$15\times15$ arcsec, significantly larger than the VLA beam
($6.99\times6.03$ arcsec). To compare the radio and sub-mm emission at
the same resolution (see Sect. \ref{sec:comparing-radio-sub}), we also
produced a tapered image of the radio data, weighting down the long
baselines to reach a Gaussian restoring beam similar to the SCUBA
beam. The resulting map is presented in Fig.\ref{tapered} (solid white
lines).

The AIPS task IMFIT was used to fit Gaussian components to all the
radio sources detected in the field (3 and 4 Gaussians simultaneously,
in the case of the two extended regions of radio emission in the
naturally weighted map). The radio positions obtained are represented
by square boxes in Fig.\ref{natural}. The results are listed in Table
\ref{radio_data}, together with their formal errors.

Radio source components \emph{Ra}, \emph{Rb}, \emph{Rc} and \emph{Fa}
appear to be related to the sub-mm emission. The source \emph{Fb} may
be related to the central brightest cluster galaxy (BCG) (see Table
\ref{offsets}), but there is no obvious optical/NIR counterpart for
\emph{Fc}.

We also detect two bright radio sources, \emph{Fd} and \emph{Fe}. The
latter is clearly identified with a optical/NIR counterpart, and both
radio sources are almost two orders of magnitude brighter than the
other radio sources in the field. These sources are probably not
lensed images of the same background source --- \emph{Fe} is more
compact than \emph{Fd}, even though it is brighter. The positions of
\emph{Fd} and \emph{Fe} are coincident (within the errors) with two
radio sources already reported in \cite{Stocke99} (see Table
\ref{offsets}).

\section{A lens model of the radio emission}  \label{sec:lens-model}


In order to aid our interpretation of the radio emission associated
with SMM J04542$-$0301, we have created a simple elliptical lens model
with external shear to describe the cluster lens potential. Our
analysis employs the GRAVLENS software package developed by
\citet{CK01}. The modeling strategy is described in
Sect.\ref{subsec:modeling-strategy}, and the results of the model are
discussed in Sect.\ref{subsec:results}.

\subsection{Modeling strategy} \label{subsec:modeling-strategy}

The first lens model of MS0451.6-0305 was presented in
\cite{T03}, and describes the total mass distribution of the cluster
by a singular isothermal ellipsoid. To be more sensitive to the local
mass distribution, \cite{B04} modeled the cluster core and 39 galaxy
cluster members using 40 smoothly truncated pseudo-isothermal
elliptical mass distribution profiles \citep[PIEMD, see ][]{kneib96}.

The critical curves of the best lens model found by \cite{B04} are
shown in Fig.\ref{natural}. It can be seen that the tangential
critical curve lies between the radio emission \emph{Ra1-Ra2},
\emph{Rb1-Rb2} and \emph{Rc1-Rc2}. Based on the general properties of
the lens geometry, this suggests that each of these image pairs
belongs to a group of 3 images produced by one source located close to
the caustic in the source plane. We propose the following scenario in
an attempt to understand the radio emission we observe in terms of
gravitational lensing:

\begin{enumerate}
  \item \emph{Ra2-Ra1} are fold images of a source \emph{Ra}, with an
expected counterpart image (\emph{Ra3}) close to \emph{ARC1ci}.
  \item \emph{Rb1-Rb2} are fold images of a source \emph{Rb} with an
expected counterpart image (\emph{Rb3}) close to \emph{Rc3}.
\item \emph{Rc1-Rc2-Rc3} are multiple images of a single source
  \emph{Rc}, located behind the cluster.
\end{enumerate}

To test this hypothesis, we implemented a new lens model for
MS0451.6-0305 using the GRAVLENS code. Since we are only interested in
testing the lensed nature of the radio emission, we modeled the
overall mass distribution of the cluster using a single mass profile
plus external shear (see Appendix 1 for a more detailed
description of the related formulae). Unlike the two previous models,
we choose an NFW profile \citep{NFW96} for the cluster mass
distribution that is consistent with observations \citep[see
e.g.][]{pointecouteau05, comerford06, bassino06} and predictions from
dark matter simulations.

To constrain the model, we performed a number of distinct
steps. First, the parameters of the mass model were  chosen in
order to reproduce the general shape of the critical lines determined
by \cite{B04} (see Fig.\ref{natural}).  Second, the positions and
fluxes of the ERO images (\emph{B1}, \emph{C1}, \emph{B2}, \emph{C2},
\emph{B3/C3}) were used as constraints for the first optimisation of
the input model. And third, we included the positions and fluxes of
the radio images as new constraints (following the previous
hypothesis) to re-optimise the model. The constraints are listed in
Tables \ref{radio_data} and \ref{optical_data}. The coordinates of the
cluster centre used in \cite{T03} were chosen as the origin of the
coordinate system.

The lens model obtained through this process turned out to have a
degeneracy between the mass and the scale radius. To break this
degeneracy, we used information about the concentration parameter
($\delta_{c}$) derived from $\Lambda$CDM N-body simulations. First, we
produced a set of 10 new models varying the core radius between 40 and
150 (a range that contains the core radius value of the degenerate
model). Following the formalism presented in \citet{bu01}, we
calculated the concentration parameter and the virial mass for the set
of new models (see Appendix 1 for details). We found that, while
$\delta_{c}$ varies between 3.3 and 9.6, the virial mass is always of
the order of $10^{15} M_{\odot}$. For a halo of that mass situated at
z=0.55, the toy model presented in \citet{bu01} predicts a
concentration parameter of $\delta_{c}=3.35$. So we plotted
$\delta_{c}$ versus the core radius for the set of new models,
interpolating the results in order to determine the core radius that
corresponds to $\delta_{c}=3.35$. Finally, we fixed the core radius of
the model to this value and re-optimized the remaining parameters.

\begin{table}[bh!]
\centering
\caption{\textbf{Optical/NIR constraints used in the lens model}. The
 columns show: position (RA, DEC), total flux density in K' band
 and predicted magnification $\mu$ (see also Table 1 in B04). The
 coordinates are given as offsets with respect to the cluster centre,
 RA(J2000)=04:54:10.8 and DEC(J2000)=-03:00:51.6 \cite[see][Table
 2]{T03}. }
\begin{tabular}{lccccccc}
\hline\hline
Name & RA & DEC & Flux in K' band & $\mu$ \\
    & J2000 ($\arcsec$)  & J2000 ($\arcsec$)& $\mu$Jy &  \\

\hline 
B1    & $29.95$ & $7.498 $ & $3.6\pm0.1$  & $8\pm1$  \\
B2    & $28.47$ & $-9.702$ & $1.9\pm0.1$  & $10\pm1$ \\
C1    & $29.95$ & $4     $ & $1.4\pm0.1$  & $10\pm1$ \\
C2    & $28.47$ & $-7.502$ & $0.9\pm0.1$  & $5\pm1$  \\
B3/C3 & $1.481$ & $-32.9 $ & $2.5\pm0.1$  & $5\pm1$  \\

\hline

\end{tabular}
\label{optical_data}
\end{table}

\subsection{Results} \label{subsec:results}

\begin{figure*}[ht!]
\begin{center}
\begin{tabular}{cc}
\includegraphics[width=7cm,angle=0]{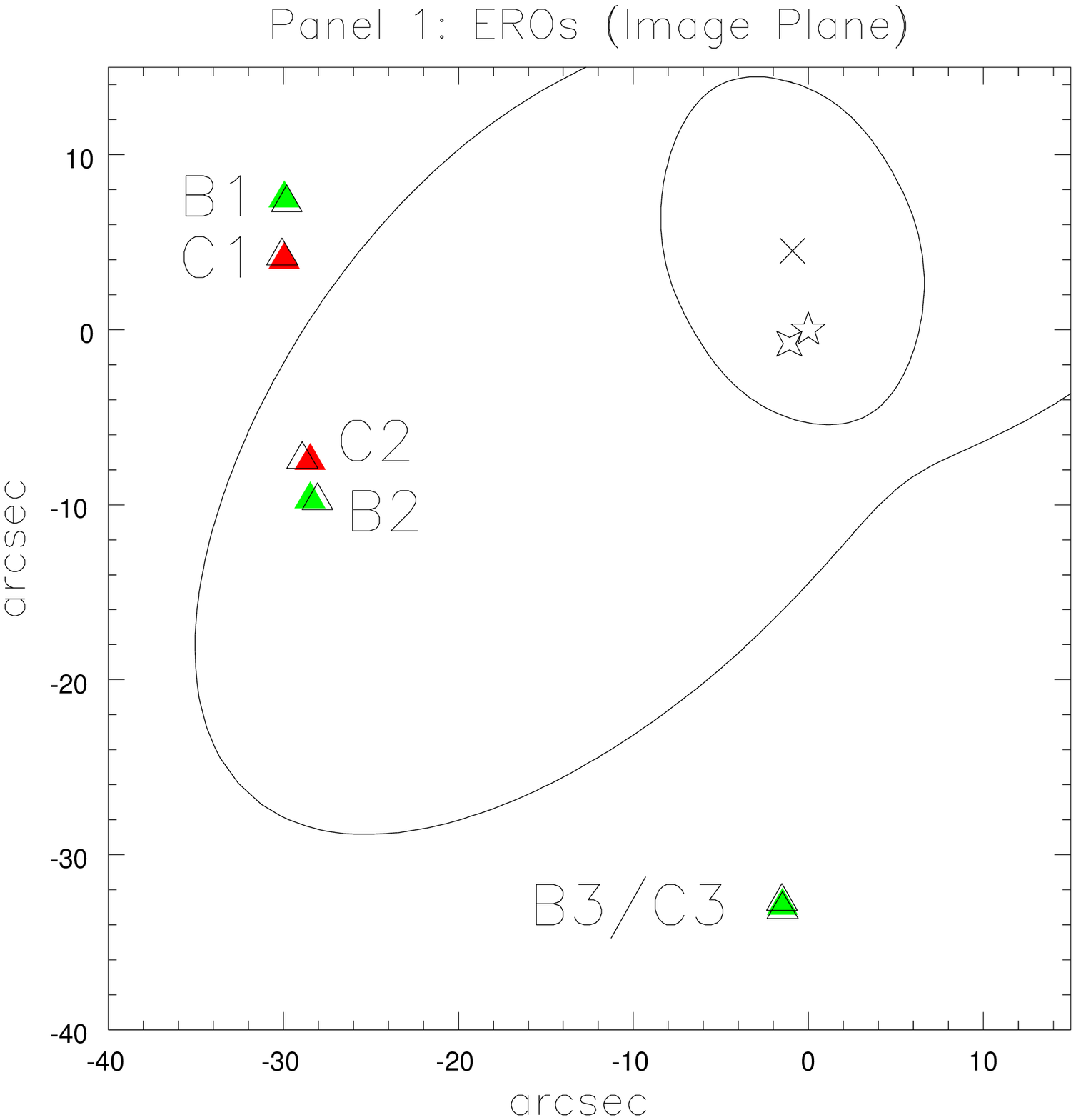} &
\includegraphics[width=7cm,angle=0]{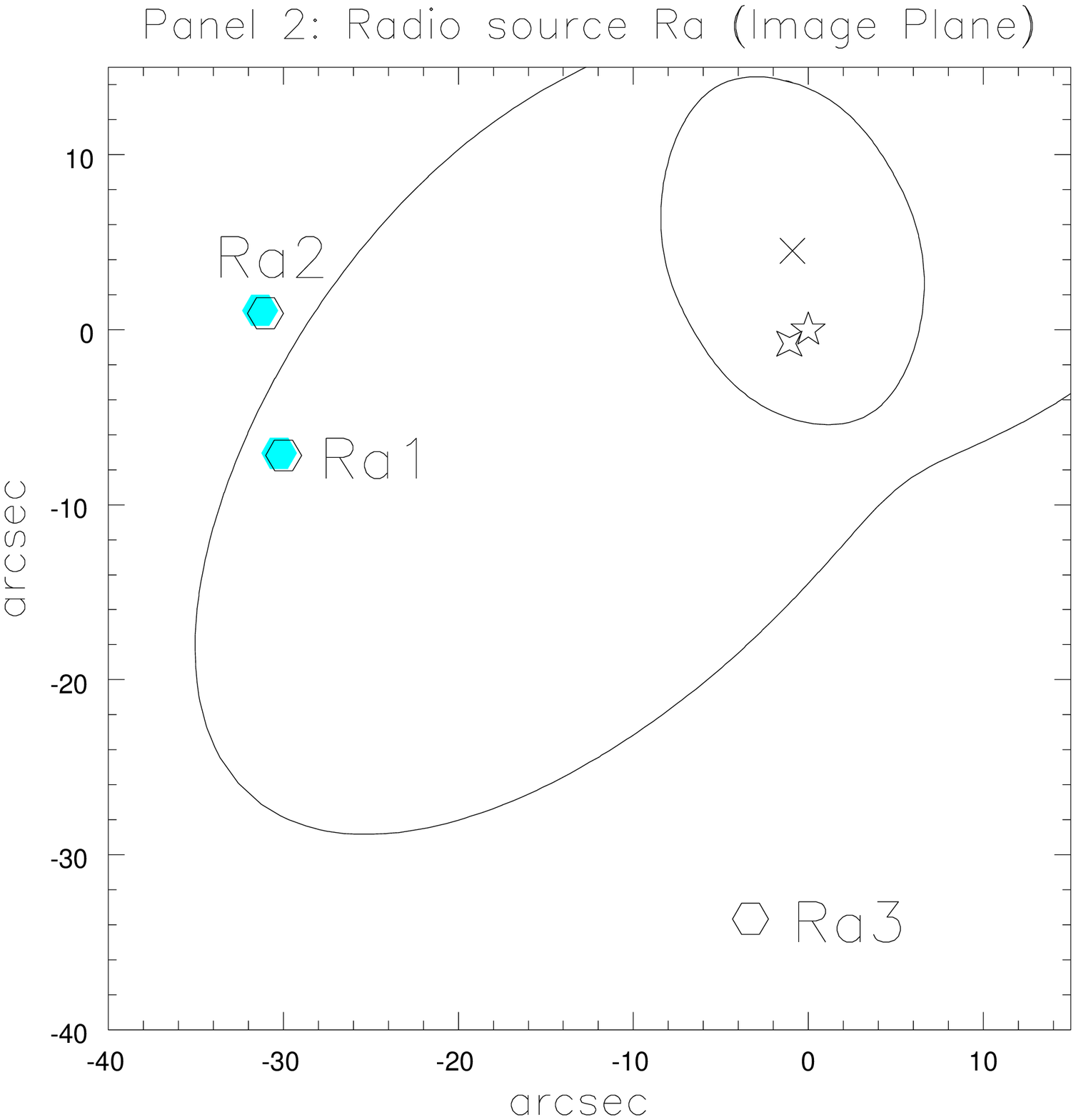} \\
\includegraphics[width=7cm,angle=0]{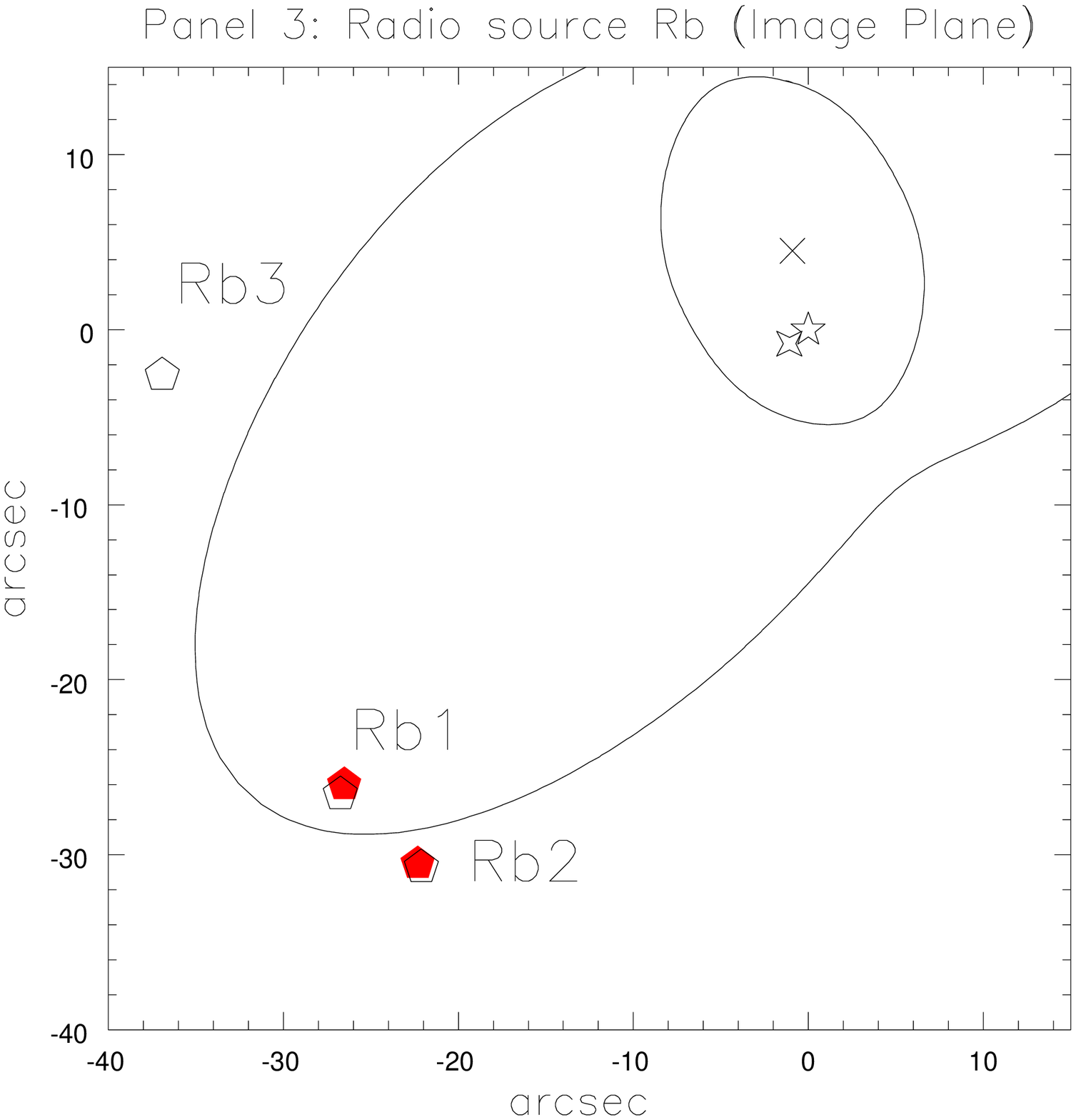} &
\includegraphics[width=7cm,angle=0]{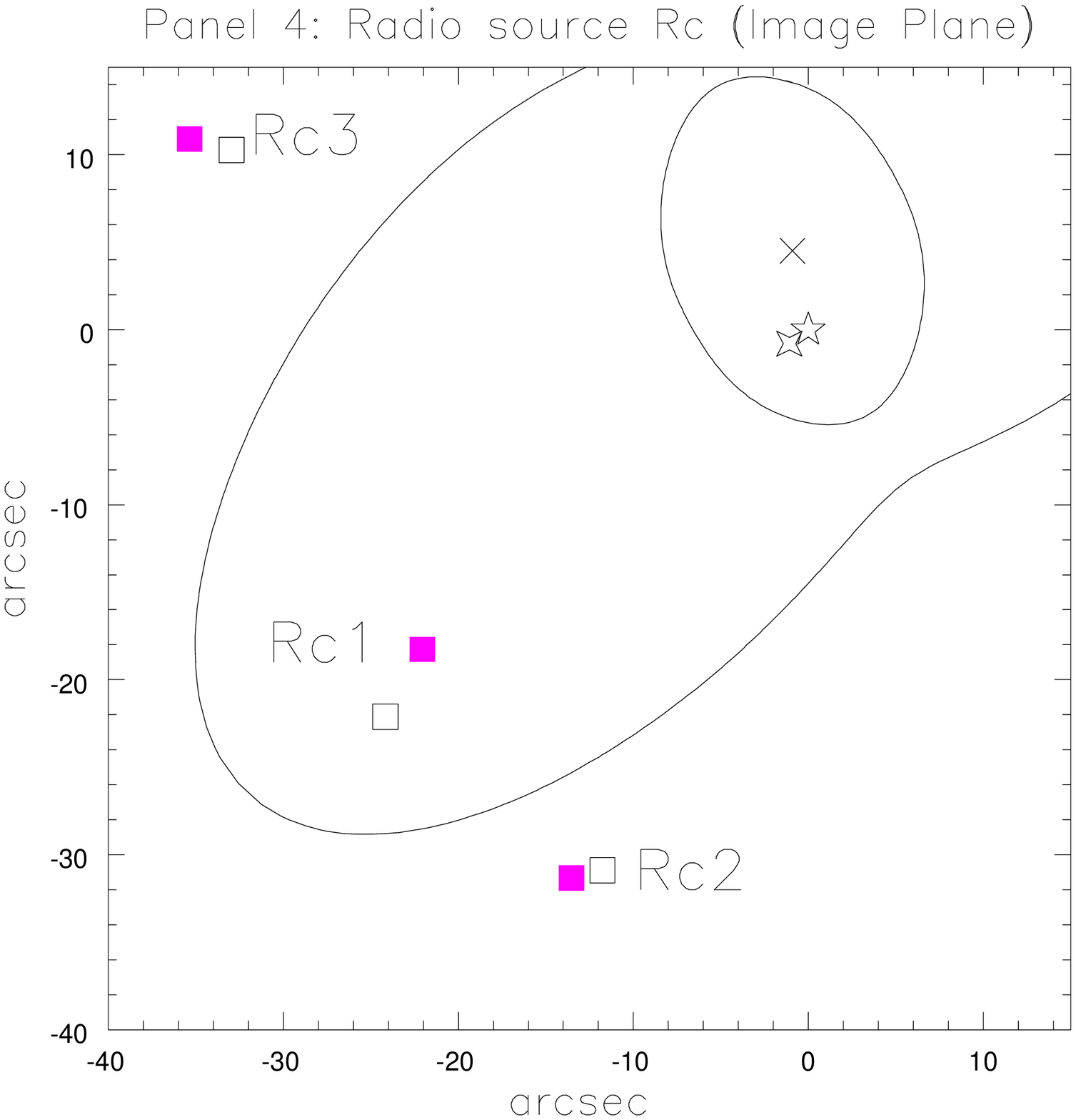} \\
\includegraphics[width=7cm,angle=0]{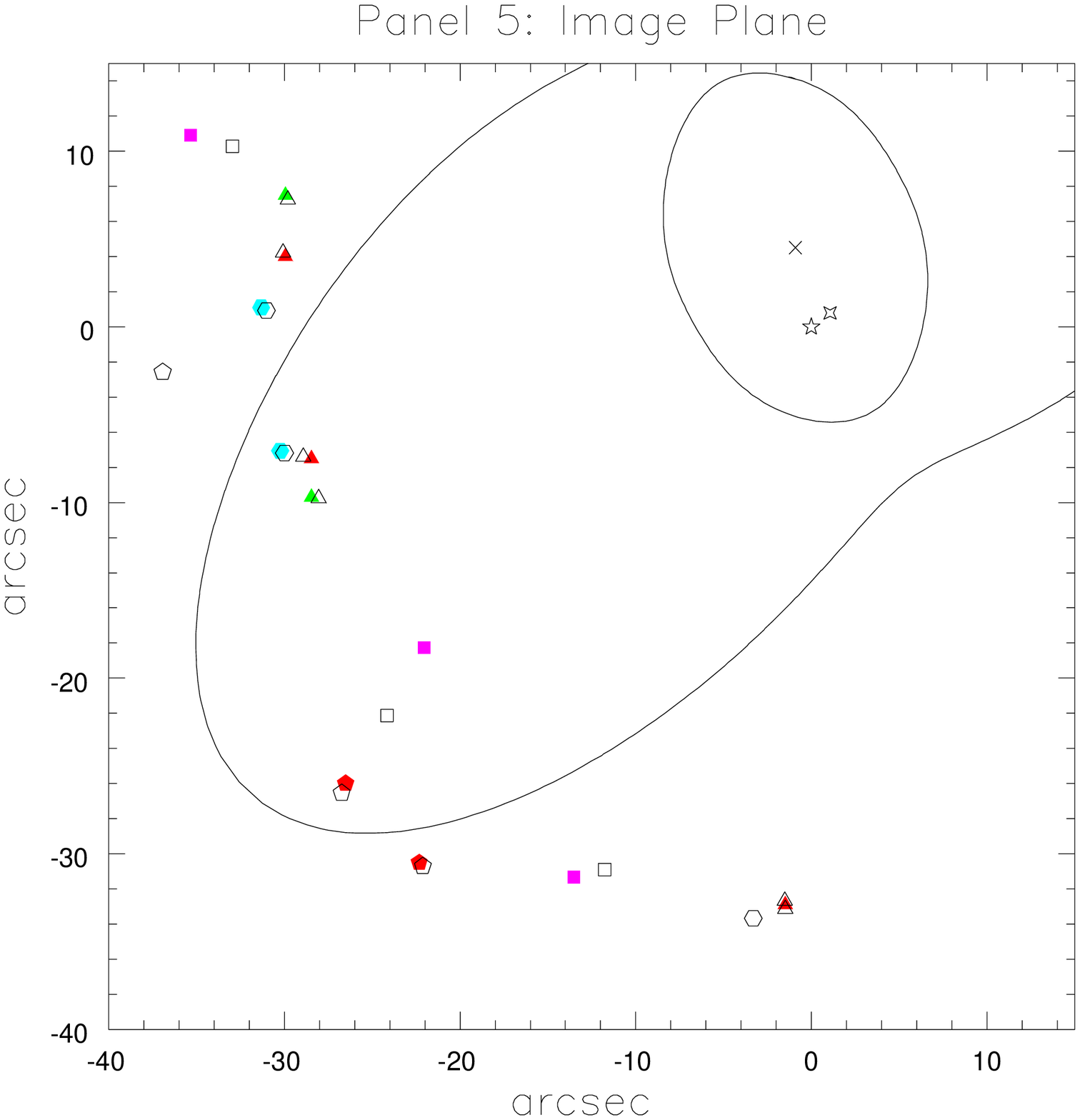} &
\includegraphics[width=7cm,angle=0]{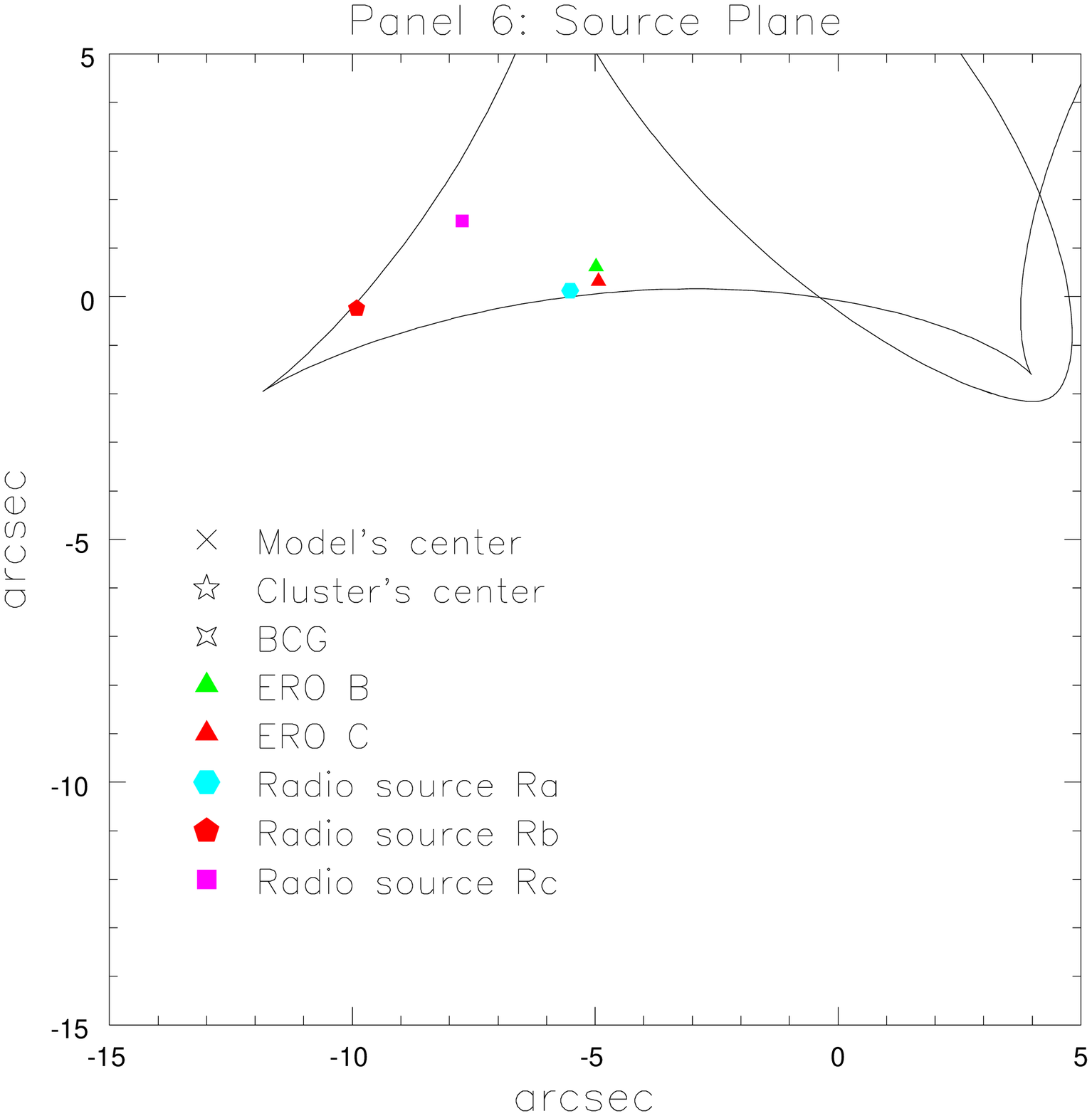} \\
\end{tabular}
\caption{\textbf{Lens model results.}  The solid black curves
represent the critical curves (panels 1 - 5) and caustics (panel 6)
that define the model. Filled symbols represent the measured positions
of the ERO images \citep{B04} and the radio components; empty symbols
are the positions predicted by the model. The ERO pair and each group
of suggested multiply-imaged radio components are shown seperately in
different panels (1 to 4). Panel 5 shows all of these components plotted
together. Panel 6 shows the position of the ERO pair and radio sources
in the source plane.}
\label{lensmodel}
\end{center}
\end{figure*}

Figure \ref{lensmodel} and Table \ref{model_errors} show the results
of the lens model. Although the model is not unique, it is able
to reproduce the positions of the ERO images and the radio emission
(\emph{Ra1}, \emph{Ra2}, \emph{Rb1}, and \emph{Rb2}) very well. The
largest offsets are found for \emph{Rc1}, \emph{Rc2} and \emph{Rc3}
(see panel 4 in Fig.\ref{lensmodel}) which are the most distant images
from the critical curves. This is consistent with the effect of
degeneracies in the global mass model near the critical curves: a
change in the model parameters produces a more significant change in
the image properties when they are located further from the critical curves.

To improve the fit of images \emph{Rc1-Rc2-Rc3}, the redshift of the
model was changed for source \emph{Rc}, but this only produces a
radial shift of their predicted positions \emph{in the same
direction}, something that cannot improve the fit shown in panel
4. Therefore, we believe that the most important contribution to this
offset is probably coming from the group of galaxies in the region
between \emph{Ra1} and \emph{Rc1} (see Fig.\ref{natural}). This group
of galaxies is expected to introduce perturbations in the overall mass
distribution of the cluster which are not accounted for in the smooth
NFW mass model.

One notable result is that the model predicts two faint counterpart
images (\emph{Ra3} and \emph{Rb3}) that do not appear in our radio
image. In our model \emph{Ra3} and \emph{Rb3} are less magnified than
\emph{Ra1-Ra2} and \emph{Rb1-Rb2} respectively. The predicted relative
magnifications suggest that \emph{Ra3} and \emph{Rb3} should appear in
our maps at the 2 and 4 $\sigma$ level respectively, but there is no
evidence for this in the radio images. We note, however, that the
predicted magnifications depend strongly on the overall mass model
employed and perturbations by individual galaxies, so they should be
treated as rough estimates of the true magnification. Therefore, the
non-detection of these images do not necessarily mean that the lens
model is wrong, since the real magnification could be less than that
predicted by this simple model.

The mass model parameters that characterise the NFW profile are
summarised in Table \ref{model_params}. The errors represent the
$1\sigma$ level of the $\chi^{2}$ function of each parameter. We note
that the shear of the model is quite large, and the model centre is
shifted $\Delta$RA=$-0.9$\arcsec and $\Delta$DEC=4.5\arcsec \ from the
assumed position of the cluster centre. These effects are most likely systematic
errors that compensate for the contribution of the group of galaxies
that we are not including in the model, and the fact that we are
forcing it to fit \emph{Rb1} and \emph{Rb2} as mirror images (a
hypothesis that we will discuss in detail in
Sect.\ref{sec:comparing-radio-sub}).

\begin{table*}[ht!]
\centering
\caption{\textbf{Lens model results}. The columns show: coordinates
of the predicted images (RA, DEC), predicted lensed flux (S$_{T}$),
offsets between the measured and predicted quantities ($\Delta$RA,
$\Delta$DEC, $\Delta$S$_{T}$) and predicted magnification ($\mu$). The
coordinates are given as offsets with respect to the cluster centre,
RA(J2000)=04:54:10.8 and DEC(J2000)=-03:00:51.6 \cite[see][table
2]{T03}.}

\begin{tabular}{lcccccccc}
\hline\hline

\\ Name & RA & $\Delta$RA & DEC & $\Delta$DEC & S$_{T}$ &
$\Delta S_{T}$ & $\mu$

\\ & ($\arcsec$) & ($\arcsec$) & ($\arcsec$) &
($\arcsec$) & $\mu$Jy & $\mu$Jy \\ 
\hline \\ 
B1 &$-29.80$  &$-0.15$  &$7.25  $ &$-2.75$  &2.82   &0.75  & $12.19 $ \\
B2 &$-28.05$  &$-0.42$  &$-9.76 $ &$0.06 $  &2.70   &0.83  & $-11.66$ \\
B3 &$ -1.50$  &$0.02 $  &$-32.66$ &$-0.24$  &1.13   &0.11  & $4.89  $ \\
C1 &$-30.07$  &$0.12 $  &$4.23  $ &$-0.23$  &1.31   &0.11  & $17.66 $ \\
C2 &$-28.92$  &$0.45 $  &$-7.40 $ &$-0.1 $  &1.26   &0.36  & $-17.00$ \\
C3 &$-1.47 $  &$-0.01$  &$-33.14$ &$0.24 $  &0.35   &0.88  & $4.75  $ \\

\hline 

Ra2 &$-31.03 $ &$-0.29$  &$ 0.94 $ &$0.17 $ &100.95 &$-5.95 $ &$ 25.18 $ \\
Ra1 &$-29.99 $ &$-0.26$  &$-7.18 $ &$0.12 $ &99.23  &$-9.77 $ &$ -24.76$ \\
Ra3 &$-3.30  $ &$-   $  &$-33.67$ &$-   $ &19.73  &$-    $ &$ 4.92  $ \\

Rb1 &$-26.75 $ &$0.23 $  &$-26.52$ &$0.54 $ &131.12 &$-19.88$ &$ -36.78$ \\
Rb2 &$-22.11 $ &$-0.22$  &$-30.70$ &$0.2  $ &111.05 &$-11.05$ &$ 31.15 $ \\
Rb3 &$-36.93 $ &$-   $  &$-2.58 $ &$-   $ &37.65  &$-    $ &$ 10.56 $ \\

Rc1 &$-24.16 $ &$2.11 $  &$-22.12$ &$3.86 $ &72.17  &$17.17 $ &$ -11.51$ \\
Rc2 &$ -11.76$ &$-1.76$  &$-30.90$ &$-0.43$ &62.34  &$-4.34 $ &$ 9.94  $ \\
Rc3 &$ -32.96$ &$-2.39$  &$10.28 $ &$0.63 $ &42.37  &$35.63 $ &$ 6.75  $ \\

%
%
%

\hline

\end{tabular}
\label{model_errors}
\end{table*}

\begin{table*}[ht!]
\centering
\caption{\textbf{Lens model parameters.} The columns show: Mass scale
($\kappa_{s}$), Galaxy position in arcsec ($x_{0}$, $y_{0}$),
ellipticity (e, $\theta_{e}$), external shear ($\gamma$,
$\theta_{\gamma}$) and scale radius in arcsec ($r_{s}$). The
errors represent the 1-$\sigma$ level of the $\chi^{2}$ fuction for each
parameter. The scale radius has no error estimations because its
value was fixed during the optimization process to be consistent with
a concentration parameter of 3.35 (see
sect.\ref{subsec:modeling-strategy})}
\begin{tabular}{lccccccccc}
\hline\hline\\ Mass Model & $\kappa_{s}$ & $x_{0}$ & $y_{0}$ & e &
$\theta_{e}$ & $\gamma$ & $\theta_{\gamma}$ & $r_{s}$ \\

\hline \\ NFW & $0.33^{+0.01}_{-0.01}$ & $-0.90^{+0.7}_{-1.2}$ & $4.51^{+0.9}_{-1.4}$ & $0.55^{+0.03}_{-0.03}$ & $-42.20^{+1.1}_{-5.6}$ &
$0.18^{+0.02}_{-0.02}$ & $61.44^{+2.4}_{-8.4}$ & [142.90] \\


\hline

\end{tabular}
\label{model_params}
\end{table*}


As already noted, this lens model is only meant to test if the
configuration of the observed radio emission can be understood as the
result of gravitational lensing.  In that sense,
Fig.\ref{lensmodel} shows that the model is able to reproduce the
position of the EROs and the radio observations reasonably well,
explaining the morphology of the radio map as the result of three
lensed background radio sources. Therefore, we expect the two EROs and
the three radio sources lie at the same redshift in the source plane
($z\sim2.9$, estimated in \cite{B04} for the EROs including the
spectroscopic redshift information of the ARC in their lens model).

But perhaps the most interesting result from the lens model is that
the source \emph{Ra} and the EROs \emph{B} and \emph{C} are predicted
to be located inside a region of about 0.6 arcsec in the source plane
(see panel 6 of Fig.\ref{lensmodel}), which corresponds to a linear
separation of only 4.7~kpc. The same situation is found in the case of the
LBG and the ERO pair, which are separated by $\sim$10 kpc in the
source plane \cite[see][]{B04}. This means that the radio source
\emph{Ra} is lying just between the LBG and the EROs, all of which are
located in a region smaller than the extent of a typical galaxy
($\sim$20 kpc). It would appear therefore, that the radio source
\emph{Ra}, the LBG and the ERO pair indeed constitute an interacting
or merging system of galaxies.

\section{Comparison of the radio, sub-mm \& optical/NIR data} \label{sec:comparing-radio-sub}

In this section we make a more detailed comparison between our radio
maps and the pre-existing sub-mm and NIR/optical data \citep{B04, T03}.

\subsection{Sub-mm vs Radio emission: morphology and flux density ratio} \label{sec:flux ratios}

\begin{figure*}[ht!]
  \centering
 
  \includegraphics[width=8cm,angle=-90]{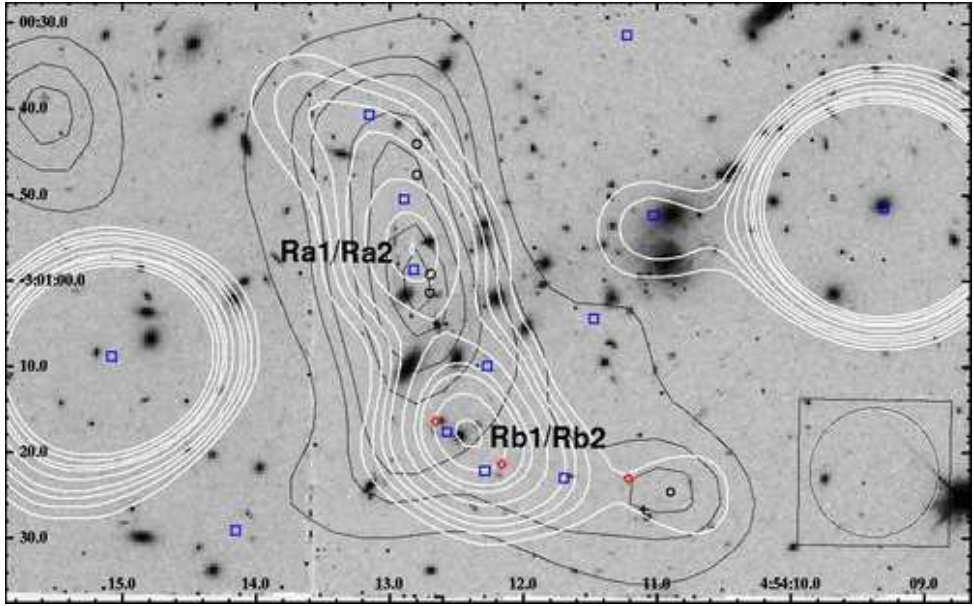}
  \caption{The VLA 1.36 GHz tapered contour map (solid white lines)
superimposed upon the SCUBA 850-$\mu$m contour map (solid black lines)
and the inverted HST F702W image of the centre of the cluster
MS0451.6-0305 \citep{B04}. The axes represent the right ascension
(x-axis) and declination (y-axis) in the J2000 coordinate
system. Contours of the \emph{tapered radio map} are drawn at -3, 3,
4, 5, 6, 8, 10, 11, 13 $\&$ 15 times the 1-$\sigma$ noise level of
14.2 $\mu$Jy/beam. Contours of the \emph{sub-mm map} are drawn at 4,
6, 7, 9, 10, 11 $\&$ 11.5 mJy/beam. The positions of the radio sources
and relevant objects in the NIR (see fig.1), are plotted as reference
points. The black circle inside the box in the bottom-right corner is
the beam size of the sub-mm map (15 $\times$ 15 arcsec), whereas the
white one (almost covered by the black circle) corresponds to the
beam-size of the radio map (15.06 $\times$ 14.26 arcsec, in position
angle $PA=68.3^{\circ}$).}.
  \label{tapered}
  \end{figure*}

Figure \ref{tapered} presents the sub-mm and tapered radio maps with a
common resolution of $\sim15\arcsec\times15\arcsec$. The regions of
radio and sub-mm emission are not only coincident, but they are
extended on the same angular scale ($\sim$ 1\arcmin) and have a
strikingly similar morphology. This strongly suggests that the radio
and sub-mm emission are associated with each other and are produced by
the same (lensed) sources, probably star forming galaxies at $z>2$. In
addition, the positions of the radio components located at \emph{Rb1},
\emph{Rb2}, \emph{Rc1}, \emph{Rc2} and \emph{Fa} (see
Fig.\ref{natural}) are consistent with the sub-mm emission that could
not be reproduced using only the ERO images and \emph{ARC1}
\citep[see][Fig.7]{B04}.

However, Fig.\ref{tapered} also shows some differences in the
morphology of the radio and sub-mm emission. The most relevant
one is that the brightest region in sub-mm is not associated with the
brightest region in radio (\emph{Rb1/Rb2}), but with the second
brightest (\emph{Ra1/Ra2}). One explanation for this apparent
discrepancy in \emph{Rb1/Rb2}, is that the radio and sub-mm emission
arise from slightly different regions in the source plane and are
differentially magnified. This effect could be quite significant for
sources lying close to or extending across a caustic. Indeed, we note
that recent Mid-IR and radio studies of local star forming galaxies
show variations across the disk of up to a factor of $\sim 5$ in the
ratio of the FIR and radio luminosity \citep{M04}. Another possibility
is that the radio emission in the region \emph{Rb1/Rb2} is not only
associated with the sub-mm emission that arises from high-z star
formation but from an additional component, perhaps an AGN in the
foreground cluster that has no counterpart in the sub-mm. Indeed, a
possible galaxy cluster member is located within 1.5\arcsec of the radio
component \emph{Rb1}, and may be an optical/NIR counterpart to
this source (see table \ref{offsets}).

The tapered map (see Fig.\ref{tapered}) also shows an
extension of the radio emission towards \emph{B3/C3} which is not seen
in Fig.\ref{natural}. This suggests the possible existence of an
extended radio source in this region, presumably associated with the
faint sub-mm emission ``toe'' that appears in the image presented by
\cite{B04}. This is consistent with the existence of the radio
counterpart Rc3 predicted by the lens model.

Assuming that the radio and sub-mm emission is produced by the same
galaxies, the $S_{850 \mu \rm m}/S_{1.4 \rm GHz}$ flux density ratio
provides information about their SEDs. Note that, since the sub-mm
image does not resolve the radio components shown in Fig.\ref{natural}
due to the poor resolution of SCUBA, the radio flux density should be
obtained from the tapered radio map. The integrated flux densities
were calculated with the AIPS task TVSTAT, using the 4 mJy/beam sub-mm
contour to delimit the same integration area in the sub-mm and tapered
radio maps (see Fig.\ref{tapered}). Using this method, we find $S_{850
\mu \rm m}=54.6 \pm 5.7 \rm mJy$ and \textbf{$S_{1.4 \rm GHz}= 0.547
\pm 0.03 \rm mJy$}. The errors were calculated using the expression
$\sigma_{beam} \times \sqrt{N}$, where $\sigma_{beam}$ is the noise
per beam of the image, and N is the number of beams within the area
delimited by the 4 mJy/beam sub-mm contour.

The observed $S_{850 \mu \rm m}/S_{1.4 \rm GHz}$ flux density ratio in
$MS0451.6-0305$ was compared with the flux density ratio obtained from
a set of SED galaxy templates. This set is composed by the archetype
star forming galaxies Arp220 and M82 (Polletta et al. in prep), the
AGN-dominated galaxy Mrk 231, and the set of Blue Compact Dwarf galaxy
SEDs presented in \cite{H05}. To detect possible differences in the
nature of faint and bright sub-mm sources, we also performed the same
analysis for the faint sub-mm source detected in A2218, and the bright
source detected in A1835 (see Table \ref{tab:fluxratios}).

We find that the observed flux density ratio in $MS0451.6-0305$ is closer to
the one obtained using the SED template of M82. However, we note that
the observed total flux ratio is being underestimated due to the
``excess'' of radio emission associated with \emph{Rb1/Rb2}, and
therefore a SED similar to Arp 220 may be more appropriate for this
source. Until the sub-mm emission can be resolved into different
components to determine their flux ratios independently, all we can
conclude is that the overall flux density ratio is largely consistent
with the favored hypothesis that the bulk of the radio/sub-mm emission
is arising from distant star forming galaxies that appear to follow
the well known FIR-radio correlation \citep{CO92, G02}.

The bright source in A1835 is the only one which is well fitted by one
of the BCD templates presented by \cite{H05}. The faint source in
A2218 seems to be similar to Mrk~231, perhaps suggesting it is a
``warm'' SCUBA source following the classification presented in \cite{E04}.

\begin{table*}
\centering
\caption{\textbf{Flux density ratios.} The columns show: name of the
sub-mm emission (SMM), redshift of the sub-mm source (z), cluster in
which the sub-mm emision is located (Cluster lens), SED template that provides the best fit to the observed flux ratio (Best SED template), flux density
ratio obtained from observations (FDR observed), flux densiy ratio
predicted by the template (FDR template), deviation between the
observed and predicted flux ratio (deviation). The value of \emph{FDR
observed} in SMM J16359+6612 was calculated by adding the fluxes of all
the images together. The deviations were calculated as
\emph{(FDR template / FDR observed)-1}.}

\begin{tabular}{ccccccccc}
\hline
\hline

 SMM         & z     & Cluster lens  & Best SED template & FDR observed         & FDR template      & deviation & References   \\
\hline						        	   
					        	   
 J14011+0252 & 2.56  & A1835         & NGC5253  & $127\pm37$   & 115       & -0.09     & I01          \\
 J16359+6612 & 2.516 & A2218         & Mrk231   & $60\pm10 $   & 49        & -0.18     & K04; G05     \\  
 J04542-0301 & 2.9   & MS0451.6-0305 & M82      & $100\pm 11$  & 115       & 0.15      & Sect. 4.1    \\

\end{tabular}
\label{tab:fluxratios}
\end{table*}

%
%

%

\begin{table*}
    \centering
    \caption{\textbf{Suggested NIR counterparts.} The columns show:
name of the radio source (Radio Source), name of the suggested NIR
counterpart (Counterpart Source), coordinates of the counterpart
source (RA cs, DEC cs), offsets between the position of the radio
source and its NIR counterpart ($\Delta$RA, $\Delta$DEC) and
references that contain information about the counterpart sources
(References). Note that $\Delta$RA, $\Delta$DEC should be interpreted
as indicative values (see Sect.\ref{subsec:look-optic-count} for
details). The coordinates are given as offsets with respect to the
cluster centre, RA(J2000)=04:54:10.8 and DEC(J2000)=-03:00:51.6
\cite[see][Table 2]{T03}. A version of this table in absolute
coordinates can be found in the online material.}

    \begin{tabular}{clccccl}
      \hline 
      \hline\\ 
      Radio Source & Counterpart Source & RA cs & DEC cs & $\Delta$RA & $\Delta$DEC & References \\ & & ($\arcsec$) & ($\arcsec$) & ($\arcsec$) & ($\arcsec$) & \\ 
      \hline \\ 
      Fb & BCG              & 1.1      & $-0.8 $ & 2.4 & 0.0  & \cite{Stocke99} \\
      Fd & 0451-03C         & 63.4     & $-16.2$ & 0.8 & 1    & \cite{Stocke99} \\
      Fe & 0451-03A         & $-21.6$  & $-1.2 $ & 0.7 & 1.1  & \cite{Stocke99} \\

      \hline

      Ra1 & ARC1 centre            & 32.2 & $-4.3 $ & 1.9    & $-2.8$  & Borys, private comunication \\
      Ra1 & ARC1 bottom end        & 31.9 & $-6.3 $ & 1.6    & $-0.8$  &  \cite{B04}, F720W HST image \\
      Ra2 & ARC1 centre            & 32.2 & $-4.3 $ & 0.8    & $-3.2$  & Borys, private comunication \\
      Ra2 & ARC1 top end           & 32.3 & $-1.7 $ & 0.9    & $2.8 $  &  \cite{B04}, F720W HST image \\
      Ra2 & B1                     & 29.8 & $-0.1 $ & $-1.4$ & $-6.4$  & \cite{B04}\\
      Ra2 & C1                     & 30.0 & $0.1  $ & $-1.4$ & $-2.9$  & \cite{B04}\\
      Ra1 & B2                     & 28.5 & $-9.7 $ & $-1.8$ & $2.1 $  & \cite{B04}\\
      Ra1 & C2                     & 28.5 & $-7.5 $ & $-1.8$ & $0.4$   & \cite{B04}\\
      Rb1 & Tc                     & 27.9 & $-24.7$ & $1.4$  & $-1.3$  & \cite{T03}\\
      Rb1 & galaxy cluster member  & 24.8 & $-25.4$ & $-1.8$ & $-0.6$  & \cite{B04}, F720W HST image\\
      Rb2 & Td                     & 20.5 & $-29.7$ & $-1.8$ & $-0.8$  & \cite{T03}\\

      \hline
   
    \end{tabular}

%
%
%
%

    \label{offsets}
 \end{table*}


\subsection{Optical/NIR counterparts to the radio emission} 
\label{subsec:look-optic-count}

\cite{B04} proposed that the sub-mm emission is related with three
objects (see Fig.\ref{natural}): A LBG (imaged as \emph{ARC1} and
\emph{ARC1ci}) and a pair of triply-imaged EROs (\emph{B}, imaged as
\emph{B1/B2/B3}; and \emph{C}, imaged as \emph{C1/C2/C3}).

As is shown in Fig.\ref{natural}, \emph{ARC1} is situated well inside
the region of radio emission associated with \emph{Ra1/Ra2}, and is
therefore probably related to it. However, in the case of the ERO
pair, the images \emph{B1} and \emph{C1} are located at the edge of
the radio emission, suggesting that they are not directly contributing
a significant amount of the radio flux density in this region.

On the other hand, as shown in Fig.\ref{natural}, although
\emph{ARC1ci} and \emph{TF} \cite[an ERO from][]{T03} are most likely
contributing to some of the measured flux density in the ``sub-mm
toe'', \emph{B3/C3} is coincident with the maximum of this
region. Therefore the EROs are certainly related to the sub-mm
emission, so we expect them to be related with the radio emission as
well.

The offsets between the estimated centres of the radio emission and
the optical/NIR candidates in the image plane are summarised in Table
\ref{offsets}. Note that the offsets of \emph{ARC1} and the ERO images
with respect to \emph{Ra1} and \emph{Ra2} are larger in declination
($\Delta\rm DEC$=3\arcsec) than in right ascension ($\Delta\rm
RA$=2\arcsec). This is probably an effect of the magnification
produced by the lens cluster, whose largest component is
preferentially aligned in the direction of declination (as is
reflected in the direction of \emph{ARC1} and the overall morphology
of the sub-mm and radio emission). In the source plane, our lens model
predicts that the offset between \emph{Ra} and the ERO pair in the
image plane is reduced to $\sim1\arcsec$. Indeed, as shown in
Sect.3.2, the LBG and the EROs probably constitute an interacting
system, with the radio source \emph{Ra} situated between them in the
source plane. This suggests that the detected radio and sub-mm
emission may come from the region in which the systems interact,
perhaps due to the enhanced star formation produced by the merging
process. A similar phenomena is also observed in the Antennae galaxy,
where the bulk of the $\lambda20$~cm radio emission is situated
between the nuclei of both galaxies \citep[see][]{Hummel86}. This
scenario can explain the offsets between the radio, sub-mm and NIR
emission observed in the region of \emph{ARC1}, \emph{B1} and
\emph{C1} in the image plane.

Moving to the southern region of Fig.\ref{natural}, the open diamonds
correspond to the positions of three additional EROs reported in
\cite{T03}. Two of these EROs (\emph{Tc} and \emph{Td}) are located
within 2 arcseconds of \emph{Rb1} and \emph{Rb2} (see table
\ref{offsets}), and may be their NIR counterparts. In this case we
expect them to be mirror images, as we assumed in our lens model.  We
note that Takata et al. argue that these EROs have different
photometric redshifts ($z_{C}=3.730$, $z_{D}=0.5$), which is
inconsistent with that hypothesis. However, both show the same colors
(within the errors) in all bands except for B and $I_{c}$, and in
those cases the differences may be due to contamination effects (from
the galaxy cluster members situated close to \emph{Tc}) and the use of
a different aperture in each source \citep[see][Table
1]{T03}. Therefore, with this information we cannot discard the
possibility that \emph{Tc} and \emph{Td} are lensed images of the same
source. On the other hand, the scenario in which \emph{Tc} and
\emph{Td} are not mirror images (but still the optical/NIR
counterparts of \emph{Rb1} and \emph{Rb2}), is also possible within
the lensing context. As is shown in Fig.\ref{natural}, \emph{Rb1} is
located very close to the critical curve presented in \cite{B04}, so a
small change in its position can move it right on top of the critical
curve (and the source component on top of the caustic), resulting in extremely high magnifications.  This is consistent with the
non-detection of the predicted counterpart image \emph{Rb3}, and the
high brightness of \emph{Rb1} in the image plane. We also note that
the shape of \emph{Tc} is extremely elongated in the same direction of
other faint arcs that appear in the same region of the Hubble image,
suggesting that \emph{Tc} may be lensed.

Another possibility (as discussed earlier in
Sect.\ref{sec:comparing-radio-sub}), is that a possible cluster galaxy
member could be the NIR counterpart of \emph{Rb1}. This scenario
can also explain the high brightness of \emph{Rb1} if the galaxy
cluster member turns out to be a ``radio loud'' AGN.

Note that four of the six optical/NIR possible counterparts of the
radio/sub-mm emission are EROs, which is consistent with the results
presented in \cite{KN05}.  On the other hand, studies carried out so
far are inconclusive with respect to the overlap between LBGs and SMGs
\citep{C02c, Adelberger98, webb02, Huang05}. However, in the scenario
proposed here, the radio and sub-mm emission is the result of an
interaction that involves a LBG, rather than emission coming from the
LBG directly.

Apart from the effects of lensing magnification, we also identify four
other possible sources of error associated with the measured offset
positions. In order of importance these include:
\begin{itemize}
\item random measurement errors in the determination of the centre of
the unresolved, blended radio components, 

\item errors in the choice of the position of \emph{ARC1}, due to its
extended and complicated structure,

\item systematic errors due to offsets between the HST and VLA
  coordinate reference systems (expected to be up to $\sim1 \arcsec$,
  but clearly not dominant since no systematic trend is shown in Table
  \ref{offsets}).

\item intrinsic offsets that can appear if the radio and sub-mm emission
come from different regions in the source plane.
\end{itemize}

Since we cannot properly estimate the contribution of these errors
(with the exception of the systematic error), the offsets
shown in Table \ref{offsets} should be taken as indicative values.

\section{Summary and Conclusions} \label{sec:conclusions}

We have presented deep VLA archive observations at 1.4 GHz of the
central region of the cluster MS0451.6$-$0305, discovering multiply-imaged
radio counterparts to the sub-mm emission SMM~J04542$-$0301, originally
discovered by \cite{C02} and recently studied by \cite{B04}. This is
the second case of multiply-lensed radio emission coming from
an intrinsically faint SMG \citep[the first case was SMM J16359+6612 in
A2218, see][]{K04, G05}. \\

With a resolution of 7 $\times$ 6 arcseconds, the radio emission
associated with SMM J0452-0301 can best be represented by seven
discrete Gaussian components. A simple lens model of this system
(based on a NFW mass profile) can reproduce the positions of the radio
components assuming that they are multiple images of 3 background
sources located at z=2.9. However, the model raises some questions
that need to be resolved.\\

Although the radio and sub-mm emission are clearly coincident and
present a similar and unusually large angular extent ($\sim$1\arcmin) and
morphology (as expected if the radio and sub-mm emission comes from
the same sources), the brightest peak of the radio emission is not
coincident with the peak in the sub-mm. We find two possible scenarios
that might explain this observation:
\begin{itemize}
\item the discrepancy is due to differential magnification produced by
  the gravitational lensing effect -- assuming that the radio and
  sub-mm emission have differnt morphologies and arise from different
  regions of the galaxy. Indeed we note, that \emph{Rb1} is situated
  very close to the critical curve presented in \cite[][see
  Fig.1]{B04}.
\item the radio emission in that region includes a component
  associated with an AGN associated with the foreground cluster. 
\end{itemize}

Borys et al. suggested that the sub-mm emission arises from an
interacting system of galaxies formed by an ERO pair and a
LBG. Although the association of the EROs with the radio emission
appears to be uncertain in the image plane (\emph{B1} and \emph{C1}
are clearly located at the very edge of the radio emission), one of
the three radio sources predicted by our lens model (\emph{Ra}) is
situated between the LBG and the ERO pair in the source plane. Our
interpretation of this result is that the interacting region of the
LBG and the ERO pair might be the source of the radio and sub-mm
emission (due to the intense star formation generated during the
merging process), whereas the optical/NIR emission might correspond to
the cores of the merging galaxies. This scenario (a situation already
observed e.g. in the Antennae galaxy) provides a consistent
explanation of the offsets between the radio, sub-mm and NIR emission
observed in the image plane (upper region of the map).\\

>From the analysis presented in \cite{B04}, it is also evident that
the LBG and the ERO pair cannot account for all the emission coming
from the central region of the sub-mm map. However, the higher
resolution VLA observations show extended radio emission located in
that region, which is expected to arise from 2 radio sources.  Two of
the components of that extended emission (\emph{Rb1} and \emph{Rb2})
seem to be related with another two EROs discovered by \cite{T03}
(\emph{Tc} and \emph{Td}).  Both EROs show similar colors, supporting
the idea (assumed in our lens model) that \emph{Rb1} and \emph{Rb2}
are images of the same source. However, this latter scenario implies
that the photometric redshifts of \emph{Tc} and \emph{Td} reported in
\cite{T03} may be incorrect.  We also found a bright galaxy (probably
a cluster member) that can be an optical/NIR counterpart of
\emph{Rb1}, keeping open the possibility that this particular region
of the radio map might include emission from a foreground AGN.  Unfortunately,
none of the evidence is compelling enough to discriminate between the
various scenarios that might
explain the nature of \emph{Rb1} and \emph{Rb2}.\\

In summary, we conclude that the radio and sub-mm emission found in
MS0451.6$-$0305 arises from at least 3 highly magnified background
sources, one of them being the interacting system proposed by
\cite{B04} (an LBG and an ERO
pair).\\

Further progress with this system requires a more complete comparison of
multi-wavelength data to be made, and a more detailed lens model to be 
constructed. Deep, mid-IR observations, as well as higher
resolution sub-mm data, might be very important in understanding this
system, in particular to confirm the possible lensed nature of
\emph{Rb1} and \emph{Rb2}. We have recently re-observed
MS0451.6$-$0305 using the VLA at 1.4 GHz in its most extended A-array
configuration. These higher resolution observations may shed new light
on this system.

\begin{acknowledgements}
The authors would like to thank Colin Borys for providing us
with the HST/SCUBA reduced images and the contours of the lens model
presented in his paper to prepare and analise our figures. We also
want to thank James Bullock for providing us his code to estimate the
concentration parameter expected for MS0451.6-0305, Charles Keeton for
helping to solve problems with the GRAVLENS code, Andy Biggs for his
assistance during the data reduction, Antonio Hern\'an Caballero and
Leslie Hunt for providing us with the SED templates, and John Stockes
for answering questions related to the bright sources in the
field. The authors are also very grateful to the referee for his/her
constructive comments which helped to greatly improve the overall
manuscript. ABA is also grateful to Edo Loenen for very valuable
comments, suggestions and help during the writing of the manuscript.
This work was supported by the European Community's Sixth Framework
Marie Curie Research Training Network Programme, Contract
No. MRTN-CT-2004-505183 ``ANGLES''.
\end{acknowledgements}

\bibliographystyle{aa}
\bibliography{5223p}

\appendix
\section{Calculation of the concentration parameter and the virial mass of the cluster}

 The NFW density profile \citep{NFW96} is defined in the
three-dimensional space \ as:

\begin{equation}
\rho(r)=\frac{\rho_{s}}{(r/r_{s})(1+r/r_{s})^{2}}
\label{rho}
\end{equation}
where $\rho_{s} $ is the characteristic density and $r_{s}$ is the
scale radius.

However the GRAVLENS code \citep{CK01} works with the projected surface mass
density of the NFW profile, which is given by:

\begin{equation}
\kappa(r)= 2 \kappa_{s} \frac{1- \mathcal{F}}{x^{2}-1} \ ; \qquad \kappa_{s}=\rho_{s} r_{s}/\Sigma_{\rm crit}
\label{kappa}
\end{equation}
where $x=r/r_{s}$, $\kappa_{s}$ is the mass scale, and $\mathcal{F}$ is
defined as:

\begin{displaymath}
\mathcal{F}(x)=\ \left\{\begin{array}{ll}
\frac{1}{\sqrt{x^{2}-1}} \tan^{-1} \sqrt{x^{2}-1} &  \textrm{($x>1$)}\\
\frac{1}{\sqrt{1-x^{2}}} \tanh^{-1} \sqrt{1-x^{2}} & \textrm{($x<1$)}\\
1 & \textrm{($x=1$)}\\
   \end{array} \right.
\end{displaymath}

Following the formalism used in \cite{bu01}, we define the scale radius as:
\begin{equation}\label{rs}
r_{s}=R_{\rm vir}/c_{\rm vir}
\end{equation}
where $c_{vir}$ is the concentration parameter and $R_{vir}$ is the virial radius.\\

Comparing the definitions of virial mass used in \cite{NFW96}
($M_{200}$) and \cite{bu01} ($M_{vir}$), the characteristic density
can be written as:
\begin{equation}
\rho_{s}=\rho_{u}(z) \ \delta_{c}
\label{rho_s} 
\end{equation}
where $\rho_{u}$(z) is the universal density at redshift z, and
$\delta_{c}$ is the characteristic over-density, which is linked with
$c_{vir}$ by the following expression:

\begin{displaymath}
\label{delta}
\begin{array}{c}
\delta_{c}=\frac{\Delta_{vir}(z)}{3}f(c_{vir}) \\
f(c_{vir})=\frac{c_{vir}^{3}}{\log(1+c_{vir})-\frac{c_{vir}}{1+c_{vir}}}\\
 \end{array}
\end{displaymath}

The parameter $\Delta_{vir}(z)$ is called the virial over-density, and
can be approximated \citep{BN98} by:
\begin{displaymath}
\begin{array}{c}
\Delta_{vir}(z) \simeq (18 \pi^{2} + 82x - 39x^{2})/ \Omega_{m}(z) - 1 \\
 x=\Omega_m(z)-1 \\
 \end{array}
\label{eq:BN98}
\end{displaymath}

Combining equations \ref{rho}, \ref{kappa}, \ref{rs} and \ref{rho_s},
the concentration parameter can be calculated using the following
expression:
\begin{equation}
f(c_{vir})=\frac{3 \ \kappa_{s} \ \Sigma_{crit}}{\rho_{u}(z_{\rm cl}) \ r_{s} \ \Delta_{vir}(\rm z_{cl})}
\label{cvir}
\end{equation}

The terms $\kappa_{s}$ and $r_{s}$ are given by the lens model (see
table \ref{model_params}). Using the estimated redshift of
MS0451.6$-$0305 from \cite{L99} ($z_{cluster}$=0.55),
$\rho_{u}(z_{cluster})$ was calculated scaling the value for z=0 given
in \cite{bu01} ($\rho_{u}(z=0)=8.3 \times 10^{10} h^{2} {\rm
M}_{\odot} {\rm Mpc}^{-3}$).

Finally, knowing the value of $c_{vir}$, the virial mass of the cluster
can then be estimated as:
\begin{equation}
M_{vir}=\frac{4 \ \pi}{3} \Delta_{vir}(\rm z_{cl}) \ \rho_{u}(\rm z_{cl}) \ 
R_{vir}^{3} 
\end{equation}

%
%

\Online
\onecolumn

\begin{table}
\centering
\caption{ \textbf{Details of the radio sources observed in the core of
 MS0451.6-0305}. The columns show: absolute coordinates in J2000
 (RA,DEC), peak flux density (S$_{Pk}$), total flux density (S$_{T}$)
 and deconvolved Gaussian sizes (major axis, minor axis and position
 angle) with their corresponding formal errors. }
\begin{tabular}{lccccccc}
\hline\hline\\
Name & RA ($+ 4^{h}$ $54^{m}$) & DEC ($- 3^{\circ}$) & S$_{Pk}$ & S$_{T}$ & Maj
Axis & Min Axis & PA \\
    & J2000 (sec)  & J2000 ($\arcmin$, $\arcsec$)& $\mu$Jy & $\mu$Jy &
$\arcsec$ &
    $\arcsec$ & deg           \\
Ra2 & $12.89\pm0.02$ & 00, $50.49\pm0.43$ &
$70\pm8$ & $95\pm18$ &
$6\pm1$ & $2\pm1$ & $27\pm28$  \\
Ra1 & $12.82\pm0.01$ & 00, $58.66\pm0.22$ &
$109\pm9$ & $109\pm9$ &
- & - & - \\
Rb1 & $12.57\pm0.01$ & 01, $17.58\pm0.16$ &
$151\pm9$ & $151\pm9$ &
- & - & - \\
Rb2 & $12.29\pm0.03$ & 01, $22.10\pm0.71$ &
$52\pm8$ & $100\pm22$ &
$9\pm2$ & $3\pm2$ & $158\pm14$ \\
Rc1 & $12.27\pm0.04$ & 01, $09.86\pm0.43$ &
$50\pm9$ & $55\pm16$ &
$6\pm2$ & - & $112\pm11$ \\
Rc2 & $11.70\pm0.04$ & 01, $22.93\pm0.81$ &
$41\pm8$ & $58\pm18$ &
$7\pm2$ & $1\pm3$ & $10\pm163$ \\
Rc3 & $13.16\pm0.05$ &  00, $40.69\pm0.46$ &
$52\pm8$  & $78\pm19$ &
$8\pm2$ & $0\pm2$ & $73\pm11$ \\
\hline
Fa & $11.47\pm0.04$ & 01, $04.34\pm0.57$ &
$45\pm9$ & $50\pm17$ &
$4\pm6$ & $0\pm4$ & $123\pm38$ \\
Fb & $11.03\pm0.04$ & 00, $52.35\pm0.59$ &
$49\pm9$ & $70\pm20$ &
$6\pm5$ & $3\pm6$ & $122\pm45$ \\
Fc & $14.16\pm0.04$ & 01, $29.06\pm0.55$ &
$44\pm9$ & $44\pm9$ &
- & - & - \\
Fd & $15.08\pm0.003$ & 01, $08.78\pm0.05$ &
$634\pm9$ & $1039\pm21$ &
$7.61\pm0.13$ & $1.36\pm0.45$ & $128\pm1$ \\
Fe & $09.31\pm0.001$ & 00, $51.53\pm0.02$ &
$1549\pm9$ & $1777\pm17$ &
$3.3\pm0.1$ & $0.8\pm0.5$ & $125\pm3$ \\
\hline
\end{tabular}
\end{table}

\begin{table}
  \begin{minipage}[t]{\columnwidth}
    \centering
    \renewcommand{\footnoterule}{} 
 \caption{\textbf{Suggested NIR counterpart sources of some radio
emissions.} The columns show: name of the radio source (Radio Source),
name of the suggested NIR counterpart (Counterpart Source), absolute
coordinates of the counterpart source in J2000 (RA cs, DEC cs),
Radio-NIR offsets ($\Delta$RA, $\Delta$DEC) and references that
contain information about the counterpart sources (References).}
    
    \begin{tabular}{clccccll}
      \hline\hline
      \\ Radio Source & Counterpart Source & RA cs ($+4^{h}$ $54^{m}$) & DEC cs ($- 3^{\circ}$) & $\Delta$RA &
      $\Delta$DEC & References \\ 
      & & ($\arcsec$) & ($\arcsec$) & ($\arcsec$) & ($\arcsec$) & \\ \hline \\

      Fb  & BCG             & 10.73  &00, 50.81   & 2.39   & 0.04     & \cite{Stocke99} \\ 
      Fd  & 0451-03C        & 15.04  &01, 08.11   & 0.75   & 0.97     & \cite{Stocke99} \\ 
      Fe  & 0451-03A        & 9.08   &00, 52.69   & 0.73   & 1.143    & \cite{Stocke99} \\

      \hline
      Ra1 & ARC1 centre     & 12.946 &00, 55.900  &0.83    & 1.9      & Borys, private comunication \\
      Ra1 & ARC1 bottom end & 12.926 &00, 57.900  & 1.62   & $-0.759$ & \cite{B04}\footnote{from F720W HST image}\\
      Ra2 & ARC1 centre     & 12.946 &00, 55.900  &$-3.19$ & 2.76     & \\
      Ra2 & ARC1 top end    & 12.953 &00, 53.280  & 0.94   & 2.79     & \cite{B04}$^{a}$\\
      Rb1 & Tc              & 12.66  &01, 16.3    & 0.54   & 1.28     & \cite{T03}      \\
      Rb1 & galaxy          & 12.453 &01, 16.98   &$-1.75$ & $-0.60$  & \cite{B04}$^{a}$\\
      Rb2 & Td              & 12.17  &01, 21.3    & 1.79   & 0.80     & Takata, private communication \\

      \hline
   
    \end{tabular}
\end{minipage}
 \end{table}

\end{document}